\documentclass[final]{aipproc}

\layoutstyle{8x11double}

\SetInternalRegister\hbadness{8000} 

\newcommand\doingARLO[2][]{%
  \ifx\mmref\undefined #1\else #2\fi
}

\begin{document}

\title
      [RR Lyrae Star]
      {Physical parameters determination of the RR Lyrae Star a CM, SW, SZ and UY in Bootes}

\classification{} \keywords{}

\author{J. H. Pe\~na}{
  address={Instituto de Astronom\'{\i}a, UNAM, Apdo. Postal 70-264, M\'exico, D.F.},
  email={jhpena@astroscu.unam.mx},
}

\iftrue
\author{A. Arellano Ferro}{
  address={Instituto de Astronom\'{\i}a, UNAM, Apdo. Postal 70-264, M\'exico, D.F.},
}

\author{L. Fox Machado}{
  address={Observatorio Astron\'omico Nacional, Apdo. Postal 877,
  22800 Ensenada, B.C.},
}

\author{M. Chow}{
  address={Observatorio Astron\'omico de la Universidad
Nacional Aut\'onoma de Nicaragua UNAN-MANAGUA},
  }

\iftrue
\author{M. Alvarez}{
  address={Observatorio Astron\'omico Nacional, Apdo. Postal 877,
  22800 Ensenada, B.C.},
}

\iftrue
\author{P. Zasche}{
  address={Instituto de Astronom\'{\i}a, UNAM, Apdo. Postal 70-264, M\'exico, D.F.},
}
\fi

\copyrightyear  {2001}

\date{\today}

\maketitle

In this preliminary work, a continuation of the determination of
physical parameters deduced from Stromgren $uvby-\beta$ photometry
for RR Lyrae stars (see Pe\~na et al., 2006, Paper I), we now
present the results obtained for the stars CM, SW, SZ, and UY in
Bootes. From the Fourier decomposition of the light curves and the
empirical calibrations developed for this type of stars the
metallicity [Fe/H] will be determined. The metallic content [Fe/H]
is also inferred from the $\Delta S$ parameter. The physical
parameters $M/M_0$, $\log (L/L_0)$, $M_V$, $\log T_{{\rm eff}}$ have
been estimated. Detailed behavior of the stars along the cycle of
pulsation has been determined from the observed photometric indices
along the pulsational phase with Stromgren $uvby-\beta$ photometry
and the synthetic indices from atmospheric models. The reddening of
the zone is found to be negligible, as was estimated in Paper I from
the reddening of several objects in the same region of the sky. The
distances to the individual objects are also estimated.

These observations were taken all at the Observatorio Astron\'omico
Nacional, M\'exico. The 1.5 m telescope to which a spectrophotometer
was attached was utilized in all seasons. The observing seasons were
several and are reported in Table 1:

\begin{table}[h]
\begin{tabular}{lllll}
\hline
  \tablehead{1}{l}{b}{Season} &
  \tablehead{1}{l}{b}{Observed star} &
  \tablehead{1}{l}{b}{Initial date} &
  \tablehead{1}{l}{b}{Final date} &
  \tablehead{1}{l}{b}{Observers\tablenote{Observers: jhp, J. H. Pe\~na; lfm, L. Fox-Machado;
  lpl, L. Parrao; ma, M. Alvarez; etl, E. Torres; mch, M. Chow; pz, P. Zasche}} \\
\hline
March 08 & whole sample &  March 25, 08 & March 28, 08 & jhp, mch, etl\\
June 09 & whole sample & June 1, 08 & June 03, 08 & jhp, lfm, ma, lpl\\
February 09 & whole sample & Feb 18, 09 & Feb 19, 09 & jhp, pz\\
\hline
\end{tabular}
\caption{ Log of the observing seasons} \label{tab:1}
\end{table}

Determination of the physical characteristics of RR Lyrae stars is
important because their accurate modeling requires knowledge of
precise metallicity plus precise determination of their physical
parameters such as effective temperature, surface gravity and
luminosity as well as an accurate determination of their periods of
pulsation. Some of these goals can be attained from the apparent
magnitude, V, and the relationships that have been developed through
the correlation between the Fourier parameters derived from the
light curves and the physical parameters of the RR Lyrae stars.
Furthermore, since we have obtained simultaneous $uvby-\beta$
photoelectric photometry of the stars, the precise variation
\underline{along the cycle} for each star can also be determined to
provide rigid constraints that the theoretical models have to
fulfill.

We have determined, as in Paper I, the reddening of different kind
of objects in the neighborhood and in the same direction of sight.
We have chosen several $\delta$ Scuti stars from Rodr\'{\i}guez et
al. (1994) in the vicinity of the RR Lyrae stars because they have
$uvby-\beta$ measurements and are next to or in the main sequence
and hence, reddening can be determined accurately. The mean
reddening is practically negligible

The variations along the cycle of pulsation can be accomplished with
the models developed for $uvby-\beta$  photometry by Lester, Gray
and Kurucz (1986, LGK86). The models have been developed taking into
account that the $uvby$ system is well designed to measure key
spectral signatures that can be used to determine basic stellar
parameters. The theoretical calibrations have the advantage of
relating the photometric indices to the effective temperature,
surface gravity, and composition. The $(b-y)$ vs. $c_1$ indices of
the stars are presented in Figure 3. It is interesting to note that
the observed star's effective temperature and surface gravity limits
as well as ranges cannot be determined with detail with only the
Fourier techniques.

\begin{table}
\begin{tabular}{llll}
\hline
    \tablehead{1}{r}{b}{Star}
  & \tablehead{1}{r}{b}{$M_v$\tablenote{From $c_0$}}
  & \tablehead{1}{r}{b}{$M_v$\tablenote{from [FeH]}}
  & \tablehead{1}{r}{b}{[Fe/H]K\tablenote{from Layden 1994}}\\
\hline
CM  & 0.442 &  0.576 &  -1.48\\
SW  & 0.641 &  0.673 &  -1.12\\
SZ  & 0.539 &  0.535 &  -1.68\\
UY  & 0.265 &  0.454 &  -2.49\\
\hline
\end{tabular}
\caption{ } \label{tab:b}
\end{table}

\begin{figure}[!b]
  \includegraphics[width=5.5cm]{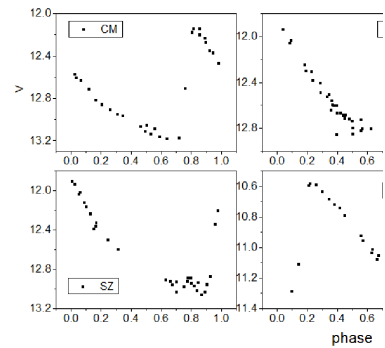}
\caption{Light curves of the observed stars.}
\end{figure}

Since RR Lyrae stars are fundamental distance calibrators, an
accurate determination of their absolute magnitudes is imperative.
Currently, we have calculated absolute magnitudes from two different
approaches. The first method employs the determined value of c1. The
second method uses the assumed metallicity. We still have to use the
morphology of the light curves. This calibration will provide us
with a more accurate metal determination of the stars. We can say
that values of the two already used methods agree, but we still have
to do the first method, one in which we have more confidence. It
would be desirable to compare these results with the derived
distances from an independent technique. Unfortunately, those
determined from Hipparcos have such large errors that a comparison
is impossible. More data will be acquired in June, 2009.

\begin{theacknowledgments}
This paper was partially supported by Papiit IN14309-3. C. Guzm\'an,
A. D\'{\i}az and F. Salas assisted us at the computing.
\end{theacknowledgments}

\end{document}